\documentstyle[twocolumn,aps,psfig,floats]{revtex}
\begin{document}
\draft
\title{Magnetic Reversal on Vicinal Surfaces}
\author{R.A. Hyman$^1$, A. Zangwill$^1$, and M.D. Stiles$^2$}

\address{$^1$School of Physics, Georgia Institute of Technology,
Atlanta, GA 30332-0430}
\address{$^2$Electron Physics Group, National Institute of Standards
and Technology, Gaithersburg, MD 20899}
\date{\today}
\maketitle
\begin{abstract}
We present a theoretical study of in-plane magnetization reversal
for vicinal ultrathin films using a one-dimensional micromagnetic
model with nearest-neighbor exchange, four-fold anisotropy at all
sites, and two-fold anisotropy at step edges. A detailed ``phase
diagram'' is presented that catalogs the possible shapes of
hysteresis loops and reversal mechanisms as a function of step
anisotropy strength and vicinal terrace length. The steps
generically nucleate magnetization reversal and pin the motion of
domain walls.  No sharp transition separates the cases of reversal
by coherent rotation and reversal by depinning of a $90^{\circ}$
domain wall from the steps. Comparison to experiment is made when
appropriate.

\end{abstract}
\section{Introduction and Background}

Laboratory studies of ultrathin films of transition metals
confirm the general principle that broken symmetry induces
magnetic anisotropy.\cite{UG,HC} The most common example is the
loss of translational invariance at the free surface of a film or
at an internal interface of a multilayer structure. The
phenomenological ``broken-bond'' model of N\'{e}el\cite{neel} then
provides an intuitive way to understand why atoms at the surface or
interface favor alignment of their magnetic moments either parallel
or perpendicular to the broken symmetry plane.\cite{CBO} In some
cases, perpendicular anisotropy occurs that is strong enough to
overwhelm the tendency for in-plane magnetization favored by
magnetostatic shape anisotropy. This situation can be exploited for
a variety of applications and has been the subject of very thorough
experimental and theoretical work.\cite{BH}

In this paper, we focus on a related phenomenon: the magnetic
anisotropy induced by crystallographic steps on the surface of a
single crystal film. Here, it is the loss of translational invariance
in directions parallel to the (nominal) surface plane that is
germane. Application of the N\'{e}el model suggests that local moments
will tend to align themselves either parallel or perpendicular to the
local step orientation. The magnitude of the effect (on a per atom
basis) is predicted to be comparable to conventional surface
anisotropy. However, it was not until 1987 that Hillebrands {\it et
al.} invoked step-induced anisotropy to rationalize their surface spin
wave data for epitaxial Fe/W(110).\cite{HBG} Since all ultrathin films
invariably have step edges (associated either with steps on the
substrate or with the nucleation and growth of monolayer height
islands during the growth process) it is not surprising that
subsequent experimental studies often cite this phenomenon in
connection with ``surface roughness effects''.\cite{rough}

We recently presented a theoretical study of in-plane magnetization
reversal in ultrathin films with step structure in typical
samples.\cite{Moschel} The model film was comprised of an array of
square, monolayer-height, magnetic islands of variable size and
density on top of a few complete magnetic layers. Classical XY-type
spins at each site were presumed to rotate in the surface plane
subject to nearest-neighbor ferromagnetic exchange, an intrinsic
four-fold in-plane anisotropy at all surface sites, Zeeman energy
from an external field, and a two-fold anisotropy at island
perimeter sites only. Numerical simulations and simple geometric
scaling arguments predicted significant variations in coercivity as
a function of coverage for layer-by-layer growth at low island
nucleation densities. This result was found to be in
semi-quantitative agreement with the surface magneto-optic Kerr
effect (SMOKE) data of Buckley {\it et al.}\cite{Buckley} for the
Cu/Co/Cu(001) system. A subsequent Monte Carlo simulation study
\cite{Rikvold} of coercivity in islanded Fe sesquilayers on W(110)
using an in-plane Ising-type spin model yielded similarly good
results in comparison to experiment.

The theoretical results of Ref.~\onlinecite{Moschel} were interpretable
on the basis of several qualitative concepts: (i) nucleation of
magnetization reversal at island edges; (ii) pinning of domain
walls at island edges; and (iii) fusion of nearby domains.
Unfortunately, even the simple island morphology studied there was
still too complex to permit a detailed analytic treatment of the
reversal process as one might desire. For this reason, we analyze
an even simpler problem in this paper: zero-temperature, in-plane
magnetization reversal in ultrathin vicinal films. The basic model
sketched above remains unchanged except that the morphology is
simplified to a periodic array of flat magnetic terraces separated
by straight, monolayer-height steps. This renders the problem
one-dimensional and amenable to analytic study.

One-dimensional models of magnetization reversal with inhomogeneous or
competing anisotropies have been a fixture of the magnetism literature
for many years. Most of these papers focus on the demonstration that
planar defects in bulk ferromagnets can nucleate reversal and/or pin
domain wall motion.  If operative, these effects call into question
the suitability of the popular single-domain, coherent rotation model
of Stoner and Wohlfarth \cite{SW} as a description of magnetization
reversal. Filipov \cite{filipov} and later Brown\cite{brown} studied
the effect of surface anisotropies on the nucleation field (where the
magnetization first deviates from its saturation value) while Mitsek
and Semyannikov\cite{MS} and later Friedberg and Paul \cite{FP}
focused on the depinning of pre-existing reversed domains as a
determinant of the coercive field (where the magnetization projected
on the external field direction first falls to zero). In recent years,
Arrott has been explicit in the application of these ideas to
ultrathin films with and without step structure.\cite{arrott} Our
analysis will be seen to substantially extend all of these studies.

On the experimental side, Heinrich {\it et al.}\cite{Heinrich}
first drew attention to the fact that a step-induced uniaxial
anisotropy must be present on vicinal surfaces. Subsequent work
confirmed this observation\cite{oepen,erskine,allenspach} and
revealed a number of other systematic features. As particular
motivation for the present work, we draw attention to the SMOKE
data of Kawakami {\it et al.}\cite{kawakami} obtained from Fe
films grown on stepped Ag(001) substrates. Characteristic
``split-loop'' hysteresis curves were found where the degree of
splitting varied smoothly with the degree of vicinality. The
authors interpreted their results using a single domain switching
model where the step edge anisotropy was distributed over the
entire surface. The analysis below will make clear the extent to
which this description can be regarded as reliable.

The plan of our paper is as follows. Section II is an overview that
includes (i) a discussion of the model assumptions; (ii) the
definition of important dimensionless quantities and the
presentation of a ``phase diagram'' that catalogs the possible
hysteresis loop topologies than can occur; (iii) a qualitative
discussion of the physical mechanisms of magnetization reversal
that can occur; and (iv) a preliminary comparison to relevant
experiments. Section III reports our mathematical procedures. We
define the Hamiltonian used and solve the model exactly to extract
the physics of zero-temperature reversal in the single domain and
single step limits. The intermediate case of multiple steps is
formulated and solved numerically.  Section IV is a discussion that
complements the earlier overview in light of our analytic and
numerical results. We consider the crossover between coherent
rotation and domain wall depinning, discuss relevant experiments in
more detail, and comment on various limitations and extensions of
the model. Section V summarizes our results and concludes the
paper.

\section{Overview}

\subsection{Model Assumptions}

We consider a uniformly thick ultrathin magnetic film adsorbed onto a
vicinal non-magnetic substrate. By flat, we mean that the film has no
island structure, {\it e.g.} a film grown in step flow
mode.\cite{markov} By ultrathin, we mean that there is no significant
variation in the magnetization density in the direction perpendicular
to the plane of the substrate terraces. By vicinal, we mean a sequence
of flat terraces of length $L$ separated by monoatomic height
steps. We assume perfectly straight steps so that the spin
configuration is a function only of the spatial coordinate
perpendicular to the steps. The problem is thereby reduced to a
one-dimensional classical spin chain with ferromagnetic exchange $J$.

\begin{figure}
{\hspace{1.5in}
\psfig{figure=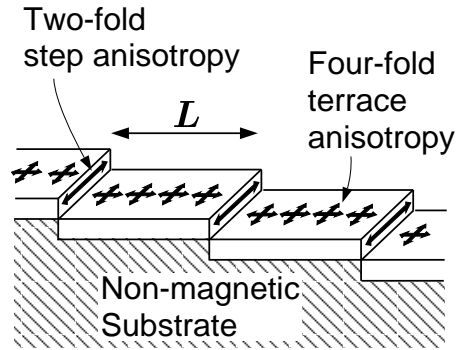,width=3.0in}}
\caption{Geometry and anisotropies for a monolayer of magnetic
material on a vicinal non-magnetic substrate.  The substrate steps
are periodically separated by a distance $L$.  There is a four-fold
anisotropy everywhere on the surface, and a strong two-fold
anisotropy localized at the steps.}
\label{fig:geom}
\end{figure}

The total surface anisotropy from all sources is presumed to compel
the spins to lie in the plane of the substrate terraces. To model
surfaces with cubic symmetry, we assign a four-fold anisotropy with
strength $K_4$ to every site of the chain and a two-fold anisotropy
with strength $K_2$ to every step site. The sign of $K_4$ is chosen to
favor spin orientations parallel and perpendicular to the
steps.\cite{rotated} If the sign of $K_2$ favors spin orientation
parallel (perpendicular) the steps, we apply the external field $H$
perpendicular (parallel) to the steps. These cases are identical by
symmetry.  Magnetostatics contributes to the total surface anisotropy
that compels the spins to lie in-plane.  For this model, with in-plane
spins, magnetostatics is not treated explicitly because its additional
effects are known to be negligible in the ultrathin
limit.\cite{magnetostatics} Figure \ref{fig:geom} is a schematic
representation of the physical situation and the spin chain model
studied here.

For simplicity, we choose units where the lattice constant $a$ is
one and $J$, $K_4$, $K_2$, and $H$, all have units of energy. To
recover dimensional units as used in
\onlinecite{Moschel}, divide $K_4$ and $K_2$ by $a^2$,
and divide $H$ by $\mu$, where $\mu$ is the atomic magnetic moment.

\subsection{The Phase Diagram}
\label{sec:pd}

\begin{figure}
{\hspace{1.5in}
\psfig{figure=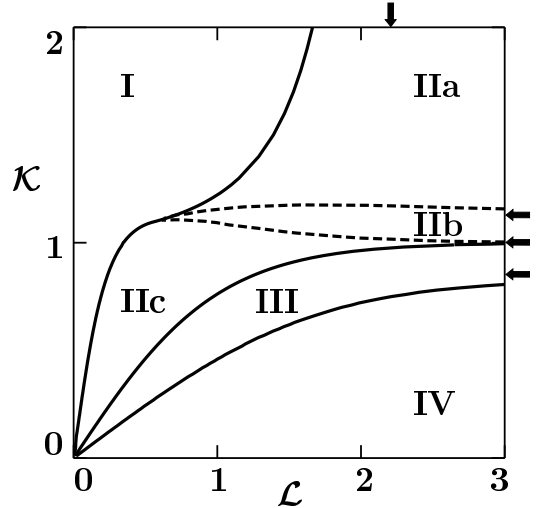,width=3.0in}}
\caption{Loop structure phase diagram.  The independent variables are
a scaled two-fold anisotropy strength at the step, ${\cal K}$ and a
scaled step separation ${\cal L}$.  Roman numerals label four
distinct loop topologies. Lower case letters label three variants
of Phase II. The vertical and horizontal arrows respectively show
the ${\cal K} \rightarrow \infty$ and ${\cal L} \rightarrow
\infty$ limits of the nearby phase boundaries.}
\label{fig:pd}
\end{figure}

We organize our discussion of hysteresis in this system around a
``phase'' diagram (Figure \ref{fig:pd}) whose axes are a
scaled step anisotropy strength ${\cal K}=K_2/2\sigma$ and a
scaled terrace length ${\cal L}=L/W$ where $K_2$ is the step
anisotropy energy, $\sigma
=\sqrt{2JK_4}$ is the domain wall energy, and $W=\sqrt{J/2K_4}$ is
the exchange length. The solid lines delineate four distinct
hysteresis loop topologies. The dashed lines divide Phase II into
three sub-variants.

\begin{figure}
{\hspace{1.5in}
\psfig{figure=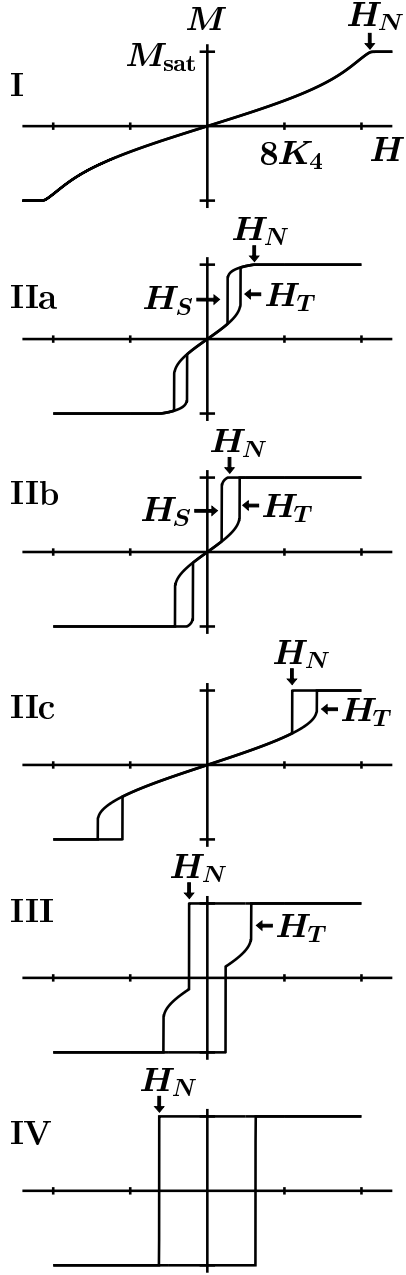,width=3.0in}}
\caption{Hysteresis loops.  The type of hysteresis loop in different
parts of the phase diagram, (see Fig.~\ref{fig:pd}).  The scaled
parameters for each loop are: I, ${\cal K}= 1.25$, ${\cal L}= 0.5$;
IIa, ${\cal K}= 1.25$, ${\cal L}= 2.0$; IIb, ${\cal K}= 1.1$,
${\cal L}= 2.0$; IIc, ${\cal K}= 0.5$, ${\cal L}=0.25 $; III,
${\cal K}= 0.5$, ${\cal L}= 0.75$; IV, ${\cal K}= 0.5$, ${\cal L}=
2.0$. }
\label{fig:loops}
\end{figure}

Figure \ref{fig:loops} illustrates representative hysteresis loops in
each phase.
Since all the loops are symmetric with respect to the sign of $H$,
it will be convenient to restrict discussion to the situation where
the field changes from positive to negative. We define three
characteristic values of the external field. The first deviation of
the magnetization from saturation occurs at the nucleation field
$H_N$. A jump in magnetization that initiates at the steps is
denoted $H_S$. A magnetization jump that initiates on the terraces
is denoted $-H_T$. $H_S=H_N$ in Phases IIc, III \& IV.

In phase I, all spins rotate continuously from the saturation
direction to the reversed direction as the external magnetic field
is reversed adiabatically. Near the left hand side of the phase
diagram, the spins rotate nearly coherently as a single unit. This
is called Stoner-Wohlfarth behavior.\cite{SW} But near the right
hand boundary of the Phase I field, the spins near the step edge
rotate more (per unit change in external field) than do the spins
near the center of the terrace. There is no hysteresis, {\it i.e.},
no jumps appear in the magnetization curve, merely more or less
spatially inhomogeneous spin rotation.

In Phase IIa, spins within an exchange length of a step rotate away
from the saturation direction at $H_N$ in response to the torque
applied by the step anisotropy. A domain wall thus forms between
the step spins and the remaining terrace spins. A field-dependent
energy barrier $\Delta_{DW}$ separates this configuration from a
configuration where all spins point nearly $90^{\circ}$ from the
saturation direction. $\Delta_{DW}
\rightarrow 0$ at $H_S$ and the domain walls ``depin'' from the steps
and sweep across the terraces. The accompanying jump in
magnetization is followed by a continuous segment of the hysteresis
curve that passes through the origin. This is an SW-like regime of
nearly coherent spin rotation. During this rotation, an energy
barrier $\Delta_{SW}$ separates the terrace spin configuration from
the nearly reversed state. At $H=-H_T$, $\Delta_{SW}$ disappears
for the terrace spins farthest from the steps and a second jump in
magnetization occurs. Reversal completes at $-H_N$ when the step
spins finally complete their rotation.

Phase IIb differs from Phase IIa because $H_T>H_N$ and the final
jump in magnetization carries the system directly to the saturated
reversed state. The phase boundary is the locus of points where
$H_T=H_N$. Note that there is a small range of ${\cal K}$ where one
encounters the phase sequence IIa $\rightarrow$ IIb $\rightarrow$
IIa as ${\cal L}$ decreases from large values.

Phase IIc mostly occupies a portion of the phase diagram where
${\cal KL} < 1$. Is this regime, the independent domain wall
description used above is no longer appropriate because the walls
have overlapped to the point where the magnetization inhomogeneity
across each terrace is not large. The reversal is better described
as nearly coherent rotation, as above, where the degree of rotation
differs for spins near and far from the steps. On the other hand, a
thin sliver of the IIc phase field extends to very large values of
${\cal L}$ where the independent domain wall picture remains valid.
This shows that there is no rigid correspondence between phases and
reversal mechanisms. More typically, as in this case, there is a
smooth crossover from a domain wall picture to a coherent rotation
picture.

Phase III occupies the smallest portion of the phase diagram. The
step anisotropy here is sufficiently small that a negative field is
needed to nucleate reversal. Otherwise, the reversal mechanism is
identical to Phase IIc.

Phase IV is characterized by $H_N<-H_T$ so that only a single
magnetization jump occurs. In fact, $H_N$ is so negative that the
state with terrace spins nearly parallel to the step is not stable
as it was in Phase III. During the jump, the degree of spatial
homogeneity of the spin rotation is dictated by the magnitude of
${\cal KL}$. Nearly coherent SW reversal occurs when ${\cal KL} \ll
1$ while rotation initiates at the step when ${\cal KL} \gg 1$.

\subsection{Relevant Experiments}

Two recent experimental studies of magnetization reversal in thin
iron films deposited onto vicinal and (nominally) flat surfaces can
be interpreted with our phase diagram. Chen and Erskine
\cite{erskine} studied ultrathin  Fe/W(001) where the step
anisotropy favors magnetic moment alignment perpendicular to the
step. Their results for an external magnetic field aligned parallel
to the steps can be compared with our results by symmetry. They
observe loops characteristic of Phase III and Phase II for the
samples they label ``smooth'' and ``stepped'' for 1.5 ML iron
coverage.

Kawakami et al.\cite{kawakami} presented a sequence of
four hysteresis loops for the Fe/Ag(001) system that we interpret
similarly as a transition from Phase III to Phase II. In this case,
the step anisotropy favors magnetic moment alignment parallel to
the step and the data they present for the external field aligned
perpendicular to the step is relevant. More details of this
comparison can be found in the Discussion section.

\section{Quantitative Analysis}

\subsection{General Results}

In the continuum limit, the model assumptions stated at the
beginning of section II lead us to the following expression for the
magnetic energy per unit length of step for an ultrathin film on a
vicinal surface:
\begin{eqnarray} \nonumber E  = \int d x \biggl[
{1 \over 2}J\left(\frac{d\theta}{d x}\right)^2 & - &{1 \over 2} K_4
\cos 4\theta
- H
\cos\theta   \\  &  + & {1 \over 2} K_2 \sum_S \delta(x - x_S)\cos
2\theta \biggr].
\label{specific}
\end{eqnarray} We remind the reader that $J$, $K_4$, $K_2$, and $H$
all have units of energy. The lattice constant is unity so the
integration variable $x$ is dimensionless. The function $\theta(x)$
is the angular deviation of the magnetization density from the
field direction at point $x$. For definiteness, we take the latter
to be perpendicular to the steps and pointing down the vicinal
staircase of Figure \ref{fig:geom}. Note that the two-fold
anisotropy acts only on step edge spins at the discrete positions
$x_S$.

We seek spin configurations $\theta(x)$ that correspond to local
minima of (\ref{specific}). In general, an energy minimum moves
smoothly in configuration space as $H$ changes and the
corresponding spin configuration and magnetization change smoothly
as well. Apart from accidental degeneracies, the only exception to
this behavior occurs when the energy minimum evolves to a saddle
point. At that point, the spin configuration changes
discontinuously, a new energy minimum is adopted, and a jump
appears in the magnetization curve. Our goal is to calculate the
field values where these jumps occur. Their number and sign
distinguish the phases of the system.

The Euler-Lagrange equation that determines the extremal
configurations of (\ref{specific}) is\cite{pend}
\begin{equation}
J{d^2\theta \over dx^2} = H \sin \theta + 2 K_4 \,\sin 4 \theta -
\sum_S \delta(x-x_S)K_2\, \sin 2\theta.
\label{EL}
\end{equation}
We seek solutions of this equation with the same periodicity as the
steps. These solutions are parameterized by two constants, the spin
angle at the center of each terrace $\theta_T$ and the spin angle
at each step $\theta_S$. One equation that relates these two is
obtained as follows. Place the origin $x=0$ at a step, multiply
(\ref{EL}) by $d\theta/dx$, and integrate from the center of the
terrace ($x=-L/2$) to an arbitrary point $x$ on the same terrace.
The result is
\begin{eqnarray}
H\cos \theta_T  +  {1 \over 2}K_4 \, \cos 4 \theta_T & = & {1 \over
2}J\Bigl({d \theta \over dx}\Bigr)^2 \nonumber\\ &+& H \cos \theta
+ {1
\over 2}K_4\, \cos 4 \theta
\label{FI}
\end{eqnarray}
using the fact that $d\theta/dx=0$ at the center of the terrace.

The constant $\theta_S$ appears  when we evaluate (\ref{FI}) at a
step. For this purpose, integrate (\ref{EL}) from $x=0^-$ to
$x=0^+$ and use reflection symmetry across the step, i.e.,
\begin{equation}
{d \theta \over dx}\biggl|_{0^+} = - {d \theta \over
dx}\biggl|_{0^-},
\end{equation}
to get
\begin{equation}
2J {d \theta \over dx}\biggl|_{0^-} = K_2\,\sin 2 \theta_S.
\label{SBC}
\end{equation}
Substitution into (\ref{FI}) yields
\begin{eqnarray} \nonumber
{\cal K}^2\,\sin^2 2 \theta_S   =  {\cal H} (\cos \theta_T & - &
\cos
\theta_S)\nonumber \\
& + & (\cos 4 \theta_T -
\cos 4 \theta_S)/2
\label{TS}
\end{eqnarray}
which relates $\theta_T$ and $\theta_S$ as desired. The scaled
magnetic field ${\cal H} = H/K_4$.

A second relation between $\theta_T$ and $\theta_S$ can be found
that involves the terrace length explicitly  by integrating
(\ref{FI}) from the center of a terrace to the step edge:
\begin{equation}
{\cal L} = 2\int_{\theta_T}^{\theta_S}{d \theta
\over
\sqrt{ [{\cal H} (\cos \theta_T - \cos \theta)+ (\cos 4 \theta_T
 - \cos 4\theta)/2]}}
\label{big}
\end{equation}

The analysis to this point is completely general and forms the
basis for all the approximate analytic and numerical results that
follow. We begin our discussion with two special situations that
can be treated in full analytically: the single domain limit and
the single step limit.

\subsection{The Single Domain limit}

This section focuses on the bottom left corner of the phase diagram
where ${\cal LK} \ll 1$. This is the Stoner-Wohlfarth limit where
only a single homogeneous magnetic domain is present. The energy
per terrace per unit length of step $\tilde{E}=E/L$ is
\begin{equation}
\tilde{E} = -\frac{1}{2}K_4\cos4\theta
+\frac{1}{2}\tilde{K}_2\cos2\theta-H\cos\theta,
\end{equation}
where the value of the effective two-fold anisotropy $\tilde{K}_2
= K_2/L$, as can be verified by substitution of a uniform spin
configuration $\theta(x)=\theta$ into (\ref{specific}).

In terms of the magnetization $M=\cos\theta$, we seek the
stationary points of the quartic expression
\begin{equation}
\tilde{E} = -K_4(2M^2-1)^2 + \tilde{K}_2M^2 -HM,
\end{equation}
i.e, the solutions of
\begin{equation}
\frac{d \tilde{E}}{d \theta}  = \sin\theta\,[H - \tilde{H}(M)]=0.
\label{note}
\end{equation}
where
\begin{equation}
\tilde{H}(M) = (2\tilde{K}_2 +8K_4)M - 16K_4M^3.
\end{equation}
The extremal condition is satisfied trivially when the
magnetization is parallel or antiparallel to the field direction
where $\sin \theta = 0$. But it is also satisfied by the cubic
equation $H=\tilde{H}(M)$. In either case, we must have
\begin{equation}
\frac{d^2 \tilde{E}}{d\theta^2} = \cos\theta\,[H - \tilde{H}(M)] +
\sin^2\theta
\frac{d \tilde{H}(M)}{d M} >0
\label{localmin}
\end{equation}
to guarantee that the solution is a local minimum of the energy.

The first term on the right hand side of (\ref{localmin})
determines the extremal properties of the $\sin \theta
=0$ solutions.
The $\theta=0$ solution is a local minimum for $H> H_N^0$ where
\begin{equation}
H_N^{0}= H_S^0 = 2\tilde{K}_2-8K_4
\label{nuc}
\end{equation}
is the limiting value of the nucleation field when ${\cal LK}
\ll 1$. Notice that portions of Phases I, IIc, III, and  IV appear in
this limit where the nucleation field $H_N^0$ and the first jump
field $H_S^0$ are coincident. The $\theta=\pi$ solution is a local
minimum for $H <
-H_N^0$.

The second term on the right hand side of (\ref{localmin})
determines the extremal properties of the $H=\tilde{H}(M)$
solutions. Because the coefficient of the cubic term is negative,
at most one of the three solutions to the cubic equation satisfies
$d
\tilde{H}(M)/d M>0$. This means that the magnetization
increases (decreases) when the field increases (decreases)--a
condition that is met when $|H| < |H_T^0|$ where
 \begin{equation} H_T^{0} = \frac{8\sqrt{6}}{9}K_4(1
+\frac{\tilde{K}_2}{4K_4})^{5/2}
\label{HT}
\end{equation}
is the limiting value of the jump field when ${\cal LK}
\ll 1$. This is true unless $\tilde{K}_2>20K_4$ in which case
the $H=\tilde{H}$ solution is stable for all values of $M$ and
there are no magnetization jumps for any value of external field.
When $\tilde{K}_2 + 8K_4<0$, the $H=\tilde{H}(M)$ solution is never
stable and the $\sin\theta
=0$ solutions are the only local minima.

The above results can be applied to find analytic formulae for the
three phase boundaries in the lower left corner of the phase
diagram. The system is in phase I when $\tilde{K}_2 > 20K_4$ since,
as noted, the magnetization curve has no jumps. The remanent slope
is $dM/dH = 1/(2\tilde{K}_2+8K_4)$. For $20 K_4>\tilde{K}_2>4K_4$,
the system is in phase IIc. The remanent slope is $dM/dH =
1/(2\tilde{K}_2+8K_4)$. For $4K_4 > \tilde{K}_2 > 2K_4$ the system
is in phase III. Phase IV occurs when $2K_4
> \tilde{K}_2$. Using these results and $\tilde{K}_2 = K_2/L$,
the boundaries between the phases near the origin are: ${\cal
K}=5{\cal L}$ between phases I and IIc, ${\cal K}={\cal L}$ between
phases IIc and III, and ${\cal K}
= \textstyle{1 \over 2} {\cal L}$ between phases III
and IV.

\subsection{The Single Step limit}

The right edge of the phase diagram where ${\cal L} \rightarrow
\infty$ is the limit where the step separation is large compared
to the exchange length and the (somewhat larger) domain wall width.
In that case, it is sufficient to study the case of a single step
bounded by semi-infinite terraces on each side. Our goal again is
to
calculate the nucleation field $H_N$ and the jump fields $H_S$ and
$H_T$. We do this by focusing attention on the spin at the step
where $\theta=\theta_S$ and the spins at $\pm \infty$ where we
assume that $\theta$ approaches the constant value $\theta_T$.

The fact that $\theta(x)
\rightarrow \theta_T$ as $x
\rightarrow
\pm
\infty$ implies that all spatial derivatives of $\theta(x)$ vanish at
infinity. Applying this to (\ref{EL}) yields
\begin{equation}
H \sin \theta_T + 2 K_4\, \sin 4 \theta_T = 0
\label{TT}
\end{equation}
which determines $\theta_T$. To find $\theta_S$, we need only note
that the $\theta_T=0$ solution to (\ref{TT}) is valid for large
values of the external field. We therefore substitute this value
into (\ref{TS}) to find
\begin{equation}
{1 \over 2}({\cal K}^2 -1 )\sin^2 2\theta_S = {\cal H}
\sin^2 \textstyle{1 \over 2} \theta_S.
\label{16}
\end{equation}
The identification
\begin{equation}
H_N^{\infty} = 8K_4({\cal K}^2 -1).
\label{HN}
\end{equation}
follows immediately since, by definition, $\theta_S$ is very small
near nucleation.  Substitution of (\ref{HN}) into (\ref{16}) gives
\begin{equation}
H=H_N^{\infty} \cos^2\theta_S \, \cos^2 \textstyle{1 \over 2}
\theta_S
\label{oops}
\end{equation}
which is valid so long as $\theta_T=0$ and $H<|H_N^{\infty}|$.

The case $H_N^{\infty}>0$ is relevant to Phases IIa and IIb where
$H_S^{\infty}$ is distinct from $H_N^{\infty}$. In particular, the
step angle $\theta_S$ increases smoothly as $H$ decreases until the
latter reaches
\begin{equation}
H_S^{\infty}=0
\label{si}
\end{equation}
when a magnetization jump occurs because (\ref{oops}) has no
solutions for $H<0$. The spin configuration just before the jump is
precisely that of a $90^{\circ}$ domain wall because
$\theta_S=\pi/2$ and $\theta_T=0$. As noted in Section II, the jump
occurs because the domain wall depins from the step and sweeps
across the terrace so that final state has $\theta(x)=\pi/2$ and
$M=0$. An explicit formula for $H_T^{\infty}$ can be found by
noting that this jump initiates with the terraces spins at $\pm
\infty$. These obey the pure Stoner-Wohlfarth dynamics of Section
II.B with $\tilde{K_2}=0$. In particular, (\ref{TT}) is identical
to (\ref{note}). The final magnetization jump thus occurs at
$-H_T^{\infty}$ where
\begin{equation}
H_T^{\infty} = {8\sqrt{6} \over 9}K_4.
\label{HTT}
\end{equation}
This value is a lower bound for the jump field when the terrace
length is finite because the presence of nearby steps retards the
final transition to the reversed state.

The case $H_N^{\infty}<0$ applies to Phases III and IV. The above
discussion shows that at nucleation in Phase III, the saturated
state jumps immediately to the spin configuration that satisfies
(\ref{TT}) with $\theta_T
\neq 0$. This state evolves smoothly until the magnetization jump
at $-H_T^{\infty}$. In Phase IV, there is only a single jump
because now (\ref{TT}) has stable solutions only at $\theta
=0$ and $\theta = \pi$ when $H=H^{\infty}_N$.

The boundaries between the various phases in the limit ${\cal L}
\rightarrow \infty$ can be found quite simply. The IIa-IIb boundary is
the locus of points where $H_T=H_N$. From (\ref{HN}) and
(\ref{HTT}) we get ${\cal K} = (1+\frac{\sqrt{6}}{9})^{1/2}\approx
1.13$. The IIb-III boundary occurs when $H_N=0$, i.e., ${\cal K}
=1$. The III-IV phase boundary is the locus of points where $H_N =
-H_T$. This gives ${\cal K} = (1-\frac{\sqrt{6}}{9})^{1/2}\approx
0.85$.

\subsection{Other Analytic Results}
This section presents three analytic results that pertain to
interior portions of the phase diagram. The first is an implicit
expression for the nucleation field at any point in the phase
diagram. The second is an exact expression for the entire boundary
between Phase IIc and Phase III. The third is the leading
correction to the Phase II jump field $H_S^{\infty}$ when the
terrace length is finite.

For the nucleation field, our interest is the first deviation of
the spin configuration from $\theta(x)\equiv0$. We thus expand
(\ref{EL}) to first order in $\theta$:
\begin{equation}
J{d^2\theta \over dx^2} = (H + 8 K_4) \theta - \delta(x)2K_2
\theta.
\label{LEL}
\end{equation}
Without the delta function, the appropriate solution to (\ref{LEL})
is
\begin{equation}
\theta = A\cosh(\sqrt{(H+8K_4)/J}x)
\end{equation}
where $A$ is a constant. Similarly linearizing the boundary
condition (\ref{SBC}) gives
\begin{equation}
2J {d \theta \over dx}\biggl|_{L/2} = -2J {d \theta \over
dx}\biggl|_{-L/2}= 2K_2\theta_S.
\label{LSBC}
\end{equation}
Combining these results yields the implicit formula
\begin{equation}
-2K_2 +  2\sqrt{J(H_N+8K_4)} \tanh \left[
\frac{L}{2}\sqrt{H_N + 8K_4 \over J}\right] =0.
\label{imp}
\end{equation}
for the nucleation field $H_N$. We obtain a more compact form by
defining a shifted and scaled nucleation field $\tilde{H}_N$ from
\begin{equation}
H_N(K_2,K_4,J,L)
=-8K_4 + 8K_4\tilde{H}_N({\cal K},{\cal L}).
\label{shift}
\end{equation}
and substituting (\ref{shift}) into (\ref{imp}). The final result
\begin{equation}
{\cal K}=\tilde{H}_N^{1/2} \tanh\left({\cal
L}{\tilde{H}_N^{1/2}}\right).
\label{scale}
\end{equation}
gives the nucleation field at any point in the phase diagram. Note
the limiting forms $\tilde{H}_N
= {\cal K}/{\cal L} $ for ${\cal K L} \rightarrow 0$ and
$\tilde{H}_N
={\cal K}^2$ for ${\cal K L} \rightarrow \infty$. These are the
Stoner-Wohlfarth and single step results obtained earlier. The line
${\cal KL} =1$ can be regarded as a crossover between the two. We
return to this point in Section IV.

The IIc-III phase boundary is defined by $H_N=0$, i.e.,
$\tilde{H}_N=1$. Substitution of this into (\ref{scale}) gives
\begin{equation}
{\cal K} = \tanh({\cal L})
\end{equation}
which is the equation of the phase boundary drawn in Figure
\ref{fig:pd}.

We turn finally to a calculation of the jump field $H_S$ in Phase
II for large but finite terrace lengths. In this limit, the domain
wall depinning picture of the jump is appropriate. The calculation
is analogous to the computation in Section III.C except that the
single step formula (\ref{TT}) is replaced by a more general
relation between $\theta_S$ and $\theta_T$ obtained from a
variational form for the spin configuration near $H_S$.

Just below (\ref{si}), we observed that the single-step spin
configuration $\theta(x)$ just before the magnetization jump at
$H_S^{\infty}$ takes the form of a $90^{\circ}$ domain wall. That
is,
\begin{equation}
\tan\theta = e^{\pm \lambda x}
\label{ex}
\end{equation}
where $\lambda = \sqrt{8K_4/J}$.  Since $\theta_S \simeq \pi/2$ at
every step, an appropriate trial function for a multi-step system
is obtained by adding together the $\pm$ wall configurations from
(\ref{ex}) in the form
\begin{equation}
\tan\theta = \tan \theta_T \cosh\lambda x,
\label{variation}
\end{equation}
which becomes
\begin{equation}
\tan\theta_S = \tan \theta_T \cosh {\cal L}
\label{var}
\end{equation}
at each step. Expanding (\ref{var}) for large ${\cal L}$ and small
$\epsilon=
\pi/2 - \theta_S$ and $\theta_T$ gives
\begin{equation}
\epsilon = \frac{1}{2\theta_T} e^{-\cal L}.
\label{eps}
\end{equation}
Performing a similar expansion on (\ref{TS}) and retaining terms to
lowest order in $H$ only yields
\begin{equation}
H = {2 \over \theta_T^2}H_N^{\infty}  e^{-2{\cal L}} + 4K_4
\theta_T^2
\label{iii}
\end{equation}
when  (\ref{eps}) is used. The jump field
\begin{equation}
H_S = 2\sqrt{8K_4 H_N^\infty}e^{-\cal L}
\label{hs}
\end{equation}
is the smallest value of $H$ for which solutions to (\ref{iii})
exist for some value of $\theta_T$.

\subsection{Numerical Results}
\label{sec:numerical}

Numerical methods were used to study three aspects of this problem:
(i) calculation of the hysteresis loops; (ii) determination of the
$\cal{L}$ dependence of the jump fields for representative values
of $\cal {K}$; and (iii) determination of the phase boundaries in
the phase diagram.

The hysteresis loops in Figure \ref{fig:loops} were computed
directly from (\ref{specific}). For each choice of control
parameters, the evolution of the stable energy minimum was followed
by a combination of conjugate gradient (CG) minimization and spin
relaxation dynamics. The initial state was chosen as the saturated
state and the external field was reversed in small steps from a
large positive value to a large negative value. The CG method
reliably follows the adiabatic minimum until a magnetization jump
occurs. But when a jump connects local energy minima that are far
separated in configuration space, the CG scheme often predicts an
obviously incorrect final state. To correct this, CG was used
consistently except in the immediate vicinity of a jump. When it
predicted a jump, the simulation  was backed up and spin relaxation
dynamics used to find the correct final state.

The nucleation field is found readily numerically from the general
formula (\ref{imp}). A more elaborate procedure is needed to find
the jump fields. Jumps in magnetization correspond to discontinuous
changes in the spin configuration. In particular, $\partial
\theta_T / \partial H$ diverges at both $H_S$ and $H_T$. But
since $\cal{L}$ is a constant for a given physical situation, it
must be the case that
\begin{equation}
{d {\cal L} \over dH} = {\partial {\cal L} \over \partial H} +
{\partial {\cal L} \over \partial \theta_T}{\partial \theta_T \over
\partial H} = 0.
\end{equation}
In this equation, ${\cal L}$ is regarded as a function of
$\theta_T$ and $H$ only since $\theta_S$ is a function of
$\theta_T$ and $H$ from (\ref{TS}). We conclude that there is a
one-to-one correspondence between the divergences of $\partial
\theta_T / \partial H$ and the zeros of $\partial {\cal L} /
\partial \theta_T$.

\begin{figure}
{\hspace{1.5in}
\psfig{figure=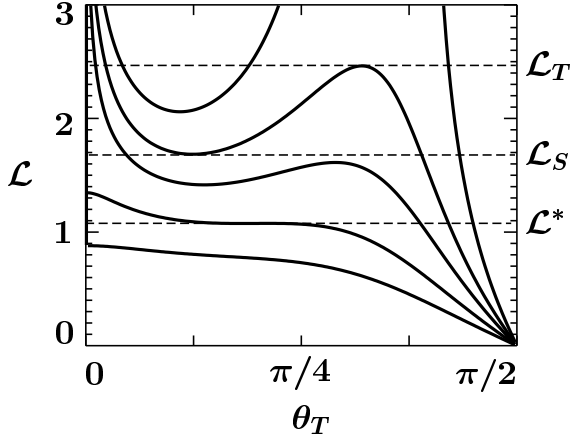,width=3.0in}}
\caption{Scaled terrace width ${\cal L}$ as a function of terrace
spin angle $\theta_T$ for ${\cal K}$= $1.25$ and ${\cal H}$=
$2,3,4,6,9$. The value of ${\cal H}$ decreases monotonically as
the sequence of curves is traversed from bottom to top. The
horizontal dashed lines labeled ${\cal L}^*$,${\cal L}_S$, and
${\cal L}_T$ are discussed in the text.}
\label{fig:l_th}
\end{figure}

The argument above directs us to find ${\cal L}(\theta_T)$ for any
desired choice of $\cal{K}$ and $\cal{H}$. Once this choice is
made, we sample many values of $\theta_T$ in the interval $0
\le \theta_T \le
\pi/2$. For each  $\theta_T$, we solve
(\ref{TS}) for $\theta_S$  and integrate (\ref{big}) to get ${\cal
L}$. Figure \ref{fig:l_th} shows ${\cal L}(\theta_T)$ for ${\cal
K}$=$1.25$ and $\cal{H}$=$2,3,4,6,9$. The value of ${\cal H}$
decreases monotonically as the sequence of curves is traversed from
bottom to top. All the curves approach either ${\cal L}
= \infty$ as $\theta_T \rightarrow 0$ or possess a semi-infinite
vertical segment at $\theta_T=0$ that begins at the point where the
curve hits the left ${\cal L}$ axis.

We now argue that the horizontal dashed line labeled ${\cal L}_S$
that is tangent to the local minimum of one of the displayed curves
defines the physical terrace width for which the corresponding
value of $H$ is exactly $H_S$. $H_S$ is encountered by reducing the
field from large positive values where the spin configuration is
saturated. The intersection of the line ${\cal L}_S$ with the
vertical portion of the curves for large ${\cal H}$ confirms that
$\theta_T=0$ at saturation. As ${\cal H}$ decreases, the
corresponding curves eventually intersect the line ${\cal L}_S$ at
small non-zero values of $\theta_T$. Finally, the intersection
occurs at the local minimum of one of the curves. This is the curve
of $H_S$ because any further reduction in field leads to a
discontinuous change in $\theta_T$ to the only remaining
intersection point on the rightmost segment of the ${\cal
L}(\theta_T)$ curves.

The horizontal dashed line labeled ${\cal L}_T$ that is tangent to
the local maximum of one of the curves defines the physical terrace
width for which the corresponding value of $H$ is exactly $H_T$.
But since Fig. \ref{fig:l_th} is drawn for $H>0$ only, the jump at
$H_T$ is encountered by increasing the external field from $H=0$
where $M=0$\cite{phaseIII}. The intersection of the line ${\cal
L}_T$ with the lowest field curve shown confirms that $\theta_T
\simeq \pi/2$. As ${\cal H}$ increases, the curves develop a local
maximum and the intersection eventually occurs at this point. This
is the curve of $H_T$ because any further increase in field leads
to a discontinuous change in $\theta_T$ to the only remaining
intersection point on the leftmost segment of the ${\cal
L}(\theta_T)$ curves.

\begin{figure}
{\hspace{1.5in}
\psfig{figure=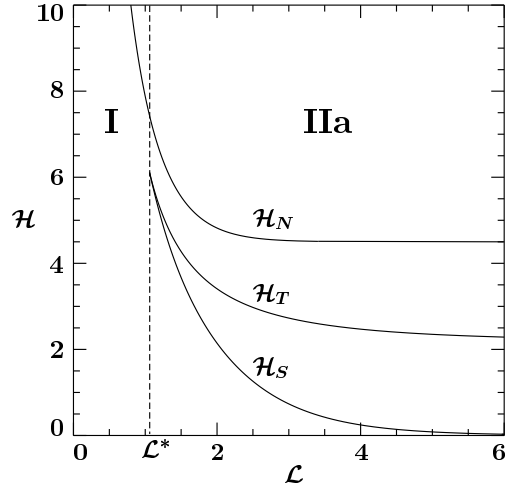,width=3.0in}}
\caption{Characteristic fields for ${\cal K}=1.25$. The vertical
dashed line ${\cal L}={\cal L}^*$ is the I-IIa phase boundary. See
text for discussion.}
\label{fig:h_l_gt}
\end{figure}

\begin{figure}
{\hspace{1.5in}
\psfig{figure=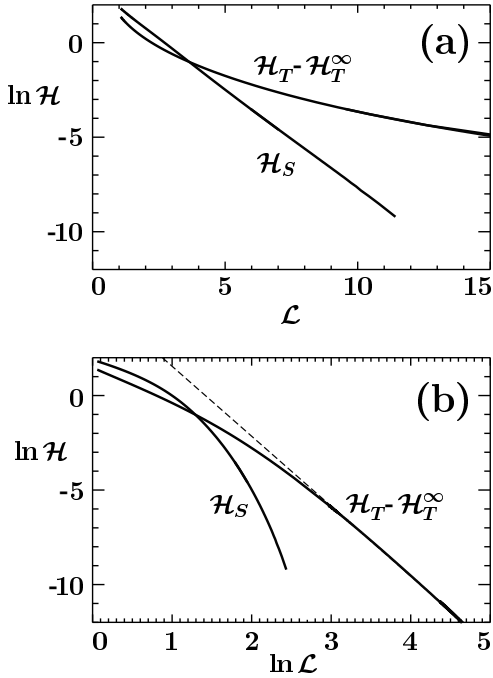,width=3.0in}}
\caption{Asymptotic behavior of the fields ${\cal H}_S$ and
${\cal H}_T-{\cal H}_T^{\infty}$ for large ${\cal L}$. Note that
$H_S^{\infty}=0$.(a) Log-linear plot.  (b) Log-Log plot. Straight
line has a slope of -3.7}
\label{fig:asymptotic}
\end{figure}

The evolution of the nucleation and jumps fields as a function of
${\cal L}$ found as described above is illustrated in Figure
\ref{fig:h_l_gt} for ${\cal K}=1.25$. Figure
\ref{fig:asymptotic}(a) confirms the exponential dependence of
$H_S$ on ${\cal L}$ predicted in (\ref{hs}). Figure
\ref{fig:asymptotic}(b) shows that $H_T \sim {\cal L}^{\chi}$ for
the last decade of data shown where $\chi
\simeq 3.7$.

The relative values of $H_N$, $H_T$, $H_S$ were used to construct
all the phase boundaries shown in Figure \ref{fig:pd}. Figure
\ref{fig:h_l_gt} is germane to the I-IIa phase boundary. No jump
fields exist for ${\cal L}<{\cal L}^*$ and $H_N > H_T > H_S$ for
${\cal L}>{\cal L}^*$. This is the same terrace length shown in
Figure \ref{fig:l_th} where the dashed line ${\cal L}={\cal L}^*$
intersects the curve of ${\cal L}(\theta_T)$ for which the extrema
(and hence the jump fields) first disappear. The I-IIa phase
boundary is asymptotically vertical as ${\cal K} \rightarrow
\infty$. The limiting value of ${\cal L}^*_{\infty}$ is found from
the same procedure as above by putting $\theta_S=\pi/2$ in
(\ref{big}).  The result is ${\cal L}^*_{\infty} \approx 2.2072$.

\begin{figure}
{\hspace{1.5in}
\psfig{figure=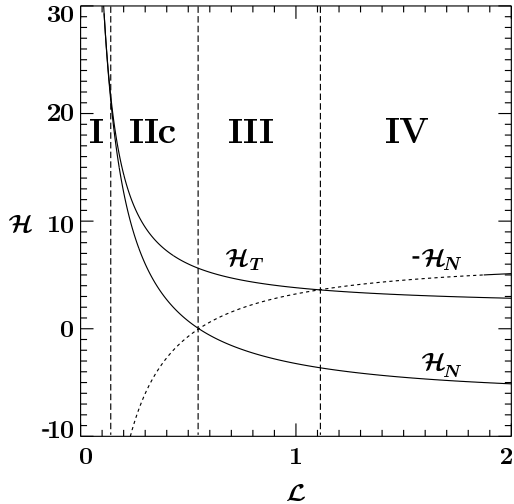,width=3.0in}}
\caption{Characteristic fields for ${\cal K}=0.5$. Vertical dashed
lines denote phase boundaries. See text for discussion.}
\label{fig:h_l_lt}
\end{figure}

\begin{figure}
{\hspace{1.5in}
\psfig{figure=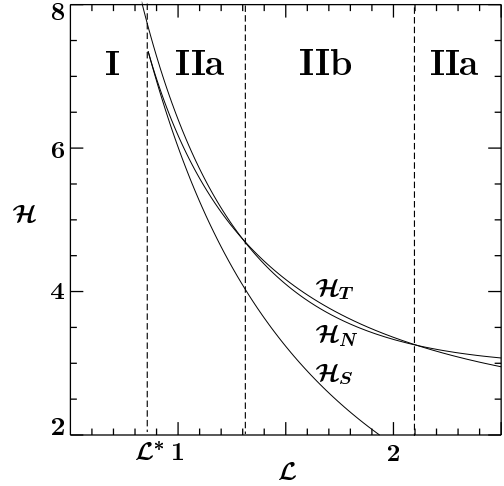,width=3.0in}}
\caption{Characteristic fields for ${\cal K}=1.17$. Vertical dashed
lines denote phase boundaries. See text for discussion.}
\label{fig:h_l_jr}
\end{figure}

Figure \ref{fig:h_l_lt} shows the ${\cal L}$ dependence of the
nucleation and jump fields for ${\cal K}= 0.5$. The absence of the
jump fields defines the range of Phase I as before. The other
phases exhibit the relative orderings of the characteristic fields
discussed in Section II, i.e.,. $H_N=0$ defines the IIc-III
boundary and $H_T=-H_N$ defines the III-IV boundary. Figure
\ref{fig:h_l_jr} shows the nucleation and jump fields for ${\cal
K}=1.17$. The re-entrant behavior IIa $\rightarrow$ IIb
$\rightarrow$ IIa described in Section II arises because the curves
of $H_T$ and $H_N$ intersect twice. The transition from IIa to IIb
at fixed ${\cal L}$ is readily understood. $H_T$ is nearly
independent of ${\cal K}$ because it is related to terrace spin
behavior far from the steps. But $H_N$ decreases rapidly as ${\cal
K}$ decreases because the torque on step spins is reduced.
Eventually, $H_N$ drops below $H_T$ for all values of ${\cal L}$.
We omit a figure that shows the IIc-IIb phase boundary ($H_S=H_N$)
explicitly.

We note finally that there is a critical point in the phase diagram
$({\cal K}_C, {\cal L}_C)$ where $H_T$, $H_S$, and $H_N$ are
coincident. This is the point in Figure \ref{fig:pd} where the
I-IIa, IIa-IIb, IIb-IIc, and IIc-I phase boundaries all meet. Our
best estimate is ${\cal K}_C\approx 1.10$ and ${\cal L}_C\approx
0.56$.

\section{Discussion}

\subsection{The Reversal Mechanism}

An important conclusion from our analysis is that a distinct
hysteresis loop topology does not imply a distinct mechanism of
magnetization reversal. This is immediately clear from Figure
\ref{fig:pd} where all four phases are present in the ${\cal LK}
\ll 1$ limit of nearly coherent rotation and three of the four phases
are present in the limit of widely separated steps where reversal
occurs by domain wall depinning. No sharp transition separates
these cases. Instead the reversal mechanism smoothly crosses over
from coherent rotation to domain wall depinning as the terrace
length or step anisotropy is increased.

The crossover is most easily understood for the case of nucleation
which, as noted, always occurs at the steps due to the torque
exerted on the saturated state by the local two-fold anisotropy.
When ${\cal LK}
\gg 1$, nucleation results in the formation of a domain of rotated
spins around each step separated from the unrotated terrace spins
by a domain wall. Now suppose that ${\cal L}$ is reduced, say, by
increasing the vicinality of the substrate. The spins on the
terrace rotate away from saturation when the domain walls begin to
overlap. In the limit when ${\cal L} \ll {\cal K}^{-1}$, their
rotation becomes indistinguishable from the rotation of the step
spins and the coherent rotation picture is a good approximation to
nucleation. Alternately, suppose that ${\cal K}$ is reduced, say,
by increasing the film thickness or by adsorbing foreign gases onto
the steps. This reduces the torque on the step spins so that their
angular deviation from the terrace spins is not as great. In the
limit when ${\cal K} \ll {\cal L}^{-1}$, this difference nearly
disappears and the coherent rotation picture is again appropriate.

We turn next the first jump field $H_S$. Coburn and co-workers
\cite{coburn} have presented a model of reversal for ultrathin
magnetic films with in-plane magnetization and four-fold
anisotropy. They assume that the film is well described by a single
homogeneous domain before and after every jump in the hysteresis
curve. Domain walls are presumed to nucleate at widely separated
surface steps or other defects. Magnetization jumps occur when the
energy density gain to make the transition $\Delta E$ is equal to a
phenomenological energy density $\epsilon$ needed to depin the wall
from the most effective pin in the film.

This description approximately reproduces our results when ${\cal
L}$ is large if we take account of the inhomogeneous spin
configuration induced by the steps. In Section II, $H_S$ was
defined as the field when the energy barrier $\Delta_{DW}$
vanished. Here, $\Delta_{DW}
=
\epsilon
-
\Delta E$ where
\begin{equation}
\Delta E \simeq -H/a + 2\sigma/L + (A/L)e^{-\cal L}
\end{equation}
and  $A$ is a constant with dimensions of energy. The first term is
the Zeeman energy gain of the saturated state compared to the
$90^{\circ}$ state. The second term is the energy cost of the
domain walls near the two steps that bound a terrace. The last term
represents an effective repulsive interaction between neighboring
walls that arises from the overlap of domain walls. The terrace
spins in the overlap region pay anisotropy energy, and the energy
of the initial state rises compared to the single step case. The
exponential dependence on wall separation is familiar from other
problems where periodic domains form, e.g, the
commensurate-incommensurate transition.\cite{bak}

The condition $\Delta_{DW}=0$ yields the estimate
\begin{equation}
H_S \simeq 2\sigma/L - \epsilon + (A/L)e^{-\cal L}.
\end{equation}
This agrees with (\ref{hs}) up to the prefactor of the exponential
if $\epsilon = 2\sigma /L$. This is not unreasonable because the
barrier for the two domain walls to depin, sweep across their
common terrace, and annihilate is associated with a spin
configuration where the two walls are separated by a distance small
compared to ${\cal L}$ but large compared to the exchange length
$W$. Of course, $\epsilon$ is not distributed across the terrace in
any physical sense. It is associated solely with the particular
spin configuration described just above.

\subsection{Comparison to Experiment}

We remarked in Section II.C that the shape of the SMOKE loops obtained
by Chen and Erskine\cite{erskine} for flat and vicinal ultrathin
Fe/W(001) appear (to the eye) to be very similar to our Phase III and
phase II topologies, respectively. To see that this is not
unreasonable, we combine the $25 {\rm \AA}$ terrace widths reported in
Ref.~\onlinecite{erskine} with typical values of the magnetic
parameters $J\sim 10^{-21}$ J, $K_2 \sim 1$ mJ/m$^{\rm 2}$, and $K_4
\sim 10^{-2}$ mJ/m$^{\rm 2}$\cite{HC} to discover that this
experiment corresponds to ${\cal L} \sim 1$ and ${\cal K} \sim 1$.
This is indeed in the vicinity of the II-III phase boundary.

We assigned the same transition to the data of Kawakami {\it et
al}.\cite{kawakami} for 25 ML of Fe on a sequence of surfaces vicinal
to Ag(001). This is still nominally an ultrathin film because the
exchange length $W
= \sqrt{J/2K_4} \sim 20$ ML using the values above. In fact, the results
of this experiment lay even closer to the lower left corner of our
phase diagram than the Chen and Erskine experiment because the
vicinality is greater.

The authors of Ref.~\onlinecite{kawakami} analyzed their data with a
single domain model similar to that of Coburn and co-workers
\cite{coburn} except that the step anisotropy was distributed
across the terraces and the depinning energy $\epsilon$ was set to
zero. Such a model actually yields no hysteresis at all--just a
magnetization curve with two symmetrical jumps. Magnetic parameters
were extracted from the experiment by matching this jump to the
average of what we call $H_S$ and $H_T$. In our opinion, formulae
similar to our (\ref{nuc}) and (\ref{HT}) for $H_S^0$ and $H_T^0$
should be used to analyze the large vicinality data\cite{ani} of
Ref.~\onlinecite{kawakami}.

\subsection{Extensions of the Model}

It is easy to think of extensions of the model studied here that would
render the results more directly comparable to experiment.  Probably
the most stringent assumption we make is that the magnetic film
smoothly coats the vicinal substrate. For relatively small terrace
lengths, this is possible if the deposition is performed at high
temperature so that nucleation of islands on the terraces is
suppressed and growth occurs in so-called ``step-flow''
mode.\cite{markov} Otherwise, it is necessary to take account of the
effect of these islands on the hysteresis. This was the subject of a
previous paper by us\cite{Moschel} for a square island geometry and
we can use those results to suggest the effect in the present case.

For fixed deposition conditions, island nucleation is increasingly
probable as the terrace length increases.\cite{markov} For this
reason, we focus on the right hand side of the phase diagram. The
magnetization jump at $H_S$ will be interrupted because the domain
walls depinned from the vicinal steps will not sweep completely
across the terraces. Instead, they will be repinned by the channels
between islands. This introduces additional jump structure into the
hysteresis curves and likely will alter the coercive field
significantly. We expect little change in $H_T$ but there will be
an extra magnetization jump before final reversal associated with
spins that remain pinned in the original saturation direction at
island edges perpendicular to the vicinal step edges.

The one-dimensional character of our model arises because we
assumed perfectly straight steps. This is not generally the case
because the desired step-flow growth mode itself induces a
step-wandering instability.\cite{bales} This instability will have
the effect of introducing two-fold anisotropies in a variety of
directions and a random anisotropy model (with spatially correlated
randomness) might be a suitable starting point in the limit of
large waviness.

Non-uniform terrace widths are another feature of real vicinal
surfaces that might also be treated in a more complete model. The
result is easy to guess in the pinned limit where every terrace
acts independently. Otherwise, nucleation and subsequent jumps will
occur first in regions of the film with largest step density and
eventually spread to regions of low step density.

Finally, we have ignored both perpendicular variations in the
magnetization and all explicit magnetostatic effects. For a vicinal
surface, dipole-dipole coupling actually induces the spins to lay
in the average surface plane of the entire crystal\cite{CBO}
rather than in the plane of the terraces as we have assumed. When
combined with crystallographic surface anisotropy, this effect
induces a two-fold anisotropy parallel to the steps at all terrace
sites.\cite{kawakami} Such a term is easily included in our basic
energy expression (\ref{specific}) and does not appreciably
complicate the analysis.

\section{Summary}
This work was motivated by the increasing awareness that the step
structure of ultrathin magnetic films can have a profound effect on
magnetic reversal and hysteresis. Our theoretical study focused on
perhaps the simplest case: a film deposited on a vicinal surface
comprised of uniform length terraces separated by monoatomic steps.
The magnetization was assumed to lay in the plane parallel to the
terraces and to vary negligibly in the direction perpendicular to
the terraces and parallel to the steps. We assumed the presence of
an intrinsic four-fold in-plane anisotropy at every site and a
two-fold anisotropy at step sites only. Explicit magnetostatics was
ignored. Attention was directed to the interesting case where one
orients an external field perpendicular to the direction of the
two-fold axes. The final model studied was a one-dimensional,
ferromagnetic spin chain in an external field with spatially
inhomogeneous anisotropy.

The analysis was performed in the continuum (micromagnetic) limit
where the spin configuration is represented by a function
$\theta(x)$ that encodes the angular deviation of the magnetization
from the external field direction. Four characteristic hysteresis
loop topologies were found and designated as ``phases'' in a
two-dimensional diagram labeled by the natural control parameters
of the model: a scaled terrace length ${\cal L}$ and a scaled step
anisotropy strength ${\cal K}$.

The hysteresis loops were characterized by a nucleation field
$H_N$, where the magnetization first deviates from saturation, a
step jump field $H_S$ where a jump in magnetization occurs from
near saturation to a state where many spins are aligned parallel to
the steps, and a terrace field $H_T$ where a jump in magnetization
occurs to the nearly reversed state. For large values of ${\cal L}$
we found $H_S \sim \exp(-{\cal L})$ and $H_T \sim {\cal L}^{-\chi}$
with $\chi \simeq 3.7$.

In all cases, reversal initiates at the steps because the torque
applied by the local anisotropy is maximal there in the saturated
state. No sharp transition separates the cases of subsequent spin
rotation by nearly coherent rotation and subsequent spin rotation
by depinning of a $90^{\circ}$ domain wall from the steps. It is a
crossover phenomenon. The coherent rotation model of Stoner
\& Wohlfarth (SW) is most appropriate in the lower left corner of
our phase diagram. The step depinning picture is most appropriate
in the upper right corner of the diagram.

To our knowledge, all existing measurements of the magnetic
properties of ultrathin films on vicinal surfaces have been
confined to a relatively small portion of our phase diagram. We
encourage experiments designed to explore the remaining {\it terra
incognita}.

\section{Acknowledgments}
The authors acknowledge an intellectual debt to Tony Arrott for the
the basic premises of this study and they are grateful to Lei-Han
Tang for his contribution to our understanding of the single step
limit. Ross Hyman was supported by National Science Foundation
Grant No. DMR-9531115.


\begin{references}
\bibitem{rotated} The case of $K_4$ favoring orientations
rotated by $\pi/4$ is qualitatively different and will be discussed
elsewhere.
\bibitem{magnetostatics} Arrott\cite{arrott} argues persuasively that
magnetostatics does not influence the hysteresis of ultrathin films
with in-plane magnetization. The present authors have confirmed
this explicitly for a multilayer version of the island model
studied in Ref.~\onlinecite{Moschel} using the computational methodology
of Mansuripur\cite{Man}.
\bibitem{pend}This is the equation of motion for a kicked rotor in an
angle-dependent potential if the variable $x$ is replaced by time.
This mechanical analogy can be helpful in much of the qualitative
analysis to follow.
\bibitem{phaseIII}This is clear for all the hysteresis loops
depicted in Figure \ref{fig:loops} where $H_T$ is labeled except
Phase III. There, there is an inner loop (not shown) that passes
through $M=0$ at $H=0$.
\bibitem{ani} Our formulae (\ref{nuc}) and (\ref{HT}) should be
modified to reflect the actual anisotropies assumed in
Ref.~\onlinecite{kawakami}.  Our point is that a pure Stoner-Wohlfarth
type analysis will be sufficient.
\end{references}
\end{document}